\documentclass[usenatbib]{mn2e}
\usepackage{amssymb}
\usepackage{times}
\usepackage{epsfig}
\usepackage{colordvi}
\newcommand{\be}{\begin{equation}}
\newcommand{\ee}{\end{equation}}

\newcommand{\ginga}{{\it Ginga}}
\newcommand{\xte}{{\it RXTE}}
\newcommand{\hete}{{\it HETE-2}}

\newcommand{\sax}{{\it BeppoSAX}}
\newcommand{\batse}{{BATSE}}
\newcommand{\ipn}{{IPN}}
\newcommand{\swift}{{\it Swift}}
\newcommand{\konus}{{\it Konus}}
\newcommand{\ulysses}{{\it Ulysses}}

\title[New soft BATSE GRBs and  spectra of X-ray rich bursts]
{New soft gamma-ray bursts in the BATSE records
and spectral properties of X-ray rich bursts}

\author[Ya.~Tikhomirova  et al.]
{Yana~Tikhomirova,$^{1}$\thanks{E-mail:
jana@anubis.asc.rssi.ru}
Boris~E.~Stern,$^{2,1}$
Alexandra~Kozyreva$^{3}$
and Juri~Poutanen$^{4}$\\
$^1$Astro Space Centre, P.N. Lebedev Physical Institute,
ulitsa Profsoyuznaya 84/32, 117997 Moscow, Russia\\
$^2$Institute for Nuclear Research, Russian Academy of Sciences,
Prospekt 60-letiya Oktyabrya 7a, 117312 Moscow, Russia \\
$^3$Sternberg Astronomical Institute, Universitetskij pr. 13, 119992 Moscow, Russia\\
$^4$Astronomy Division, P.O.Box 3000, FIN-90014 University of Oulu, Finland}

\begin{document}
\date{Accepted, Received}
\pagerange{\pageref{firstpage}--\pageref{lastpage}} \pubyear{2005}
\maketitle

\label{firstpage}

\begin{abstract}
A population of X-ray dominated gamma-ray bursts (GRBs)
 observed by \ginga, \sax\ and \hete\
should  be represented in the \batse\ data as presumably soft bursts.
We have performed a search for soft GRBs
in the \batse\ records in the 25--100 keV energy band.
A softness of a burst spectrum can be a reason why it has been
missed by the on-board procedure
and previous searches for  untriggered
GRBs tuned to 50--300 keV range.
We have found a surprisingly small number
($\sim 20$ yr$^{-1}$  down to 0.1 ph cm$^{-2}$ s$^{-1}$)
of soft GRBs where the count rate is
dominated by 25--50 keV energy channel.
This fact as well as the analysis of \hete\ and
common \sax/\batse\ GRBs indicates that
the majority of GRBs with a low $E_{\rm peak}$
has a relatively hard tail with the high-energy
power-law photon index $\beta>-3$.
An exponential cutoff in GRB spectra below 20 keV
may be a distinguishing feature separating non-GRB events.
\end{abstract}


\section{Introduction}

Observations of the prompt gamma-ray bursts (GRBs) emission
by different instruments show that
their spectra can extend from several keV up to a few MeV
\citep{whe73,tro74,met74}
sometimes up to GeV range \citep{som94}.
According to recent broadband observations by
\ginga\ \citep{str98}, \hete\ \citep{sak04}
and combined results of \sax/\batse\ \citep{kip02}
and \xte/\ipn\ \citep{smi02},   most GRBs
exhibit a peak in the $E F_{E} $ spectrum at an
energy $E_{\rm peak}$ in the 50--400 keV range. However,
distribution of $E_{\rm peak}$ is broad and large part
of events demonstrate significant X-ray (2--30 keV)
emission (X-ray dominated GRBs, X-ray rich GRBs).
At this moment study of broadband spectra
is complicated because of insufficient
statistics accumulated by broadband instruments
and biases due to different instrument responses.

The \batse\ \citep{pac99}  data of all-sky 9.1 years
(1991--2000) continuous monitoring in $\gamma$-ray range
give unique possibility for combined GRB analysis
with X-ray observations.  \batse\ $\gamma$-ray detectors
were the most sensitive instruments of this type
over GRB history. Only recently launched \swift\
experiment \citep{geh04}  has
a more sensitive $\gamma$-ray detector.
However, during the next several years \swift\
cannot accumulate statistics comparable to that of the \batse.
\batse\ detected about 2700 GRBs with fluxes down to
$\sim 0.3$ ph cm$^{-2}$ s$^{-1}$ \citep{pac99}.
In addition, the off-line scans of
the \batse\ continuous records almost doubled
the number of observed GRBs  with fluxes down to
$\sim 0.1$ ph cm$^{-2}$ s$^{-1}$
(see \citealt{kom01};
\footnote{Non-triggered supplement
to the \batse\ GRB catalogs is available at
http://space.mit.edu/BATSE/intro.html}
 \citealt{ste01}
\footnote{The uniform catalog of GRBs
found in the continuous \batse\ daily records is available at\\
http://www.astro.su.se/groups/head/grb\_archive.html})

The \batse\ detectors were sensitive to photons
from  $\sim$25 keV up to $\sim$1 MeV.
However, the on-board procedure and most off-line
searches identified GRBs according
to the signal  in the 50--300 keV range while
GRBs with a soft spectrum could be missed.
These soft events can help to outline the place
of the X-ray dominated bursts in the GRB variety.

The 25-50 keV range was inspected only in
the off-line search of \citet{kom01}.
Their scan have covered 6 out of 9.1 years
of the \batse\ data and yielded 50
 unknown low-energy events
some of which are probably soft GRBs.
Even if all of them are GRBs, the number
of these events is 50 times smaller than
that found in the 50--300 keV range.

We performed a search for GRBs,
inspecting the 25--50 keV range for time period
not covered by the scan of \citet{kom01}
with a more careful and optimized for soft GRBs
procedure.
The continuous daily 1.024 s time
resolution DISCLA records of count rate
in 8 \batse\ detectors in 4 energy channels
(25--50, 50--100, 100--300 and 300--1000 keV) were used.
We have applied the same technique and the same algorithm
as in our scan of the \batse\ DISCLA data
in the 50--300 keV range \citep{ste01}
setting the trigger in the 25--100 keV range
(i.e. in the 1st and 2nd energy channels).

\begin{table*}
\begin{minipage}{180mm}
\caption{Long ($>2$ s) GRBs found in the present scan
of the \batse\ DISCLA records (TJD~11000-11699) in 25--100 keV band.}
\begin{center}
\begin{tabular}{|rrrcrrrrrr|}
\hline
Date & Seconds & TJD & $P^a$ & HR$^b$ & $\alpha^c$ & $\delta^c$ & $R^d$&  $T_{90}^e$& $N_{50}^f$\\
     &of TJD   &   &  &  & deg & deg & deg & s & \\
 \hline\hline
\multicolumn{10}{|c|}{13 soft GRBs ($P_{25-50\ \mbox{keV}} > P_{50-300\ \mbox{keV}}$)} \\
\hline
980726& 63036& 11020&  0.11& 0.57& 255.8&  $-$54.8&  9.7&   56&   29\\
980804& 50914& 11029& 0.46& 0.87& 173.3&  $-$52.7&  7.3&   13&    5\\
980927&  6133& 11083& 0.33& 0.89&   9.6&  $-$54.5& 11.4&    6&    4\\
981225& 76754& 11172& 0.22& 0.45& 161.9&  $-$61.3& 17.0&   25&    4\\
990304& 77277& 11241& 1.85& 0.83&  31.6&  $-$26.7&  4.6&    4&    2\\
990513&  2453& 11311& 0.18& 0.30& 236.4&  $-$59.6& 16.5&   15&    2\\
990610& 20227& 11339& 0.11& 0.43& 234.8&   16.6& 17.3&   80&   11\\
990804& 39065& 11394& 0.05& 0.92&  44.1&   21.2& 36.6&   38&   10\\
990907& 75723& 11428& 0.06& 0.66& 301.0&  $-$39.3&  8.3&  126&   33\\
991003& 30847& 11454& 0.16& 0.60& 253.8&   33.2& 21.1&   13&    6\\
991009& 30691& 11460& 0.10& 0.47& 107.9&    3.5& 12.9&   24&    8\\
991106& 59880& 11488& 0.10& 0.33& 284.7&  $-$58.2& 20.5&   39&   11\\
000107&  8784& 11550& 0.12& 0.91&  74.9&  $-$61.6& 16.1&   73&   12\\
\hline
\multicolumn{10}{|c|}{8 classic GRBs ($P_{25-50\ \mbox{keV}} < P_{50-300\ \mbox{keV}}$)} \\
\hline
980707&  9097& 11001& 0.40& 3.29&  79.0&   40.4&  9.9&    6&    4\\
980930& 83166& 11086& 0.37& 1.90& 132.0&  $-$70.9&  6.7&   38&   19\\
981012& 21270& 11098& 0.11& 1.24&  59.0&   15.5& 20.7&   17&   11\\
981019& 69630& 11105& 0.29& 2.18& 208.9&  $-$40.3& 11.7&   17&    3\\
981221& 18020& 11168& 0.97& 1.82&  71.9&    3.6& 15.4&    9&    1\\
990303& 74922& 11240& 0.25& 1.00& 199.9&   52.4& 18.9&    9&    7\\
000324& 36745& 11627& 0.08& 2.10&  58.7&   26.4& 18.8&   36&   10\\
000523& 49912& 11687& 0.41& 1.01& 269.1&   80.6&  8.6&   24&    9\\
\hline
\multicolumn{10}{l}{$^a$ Peak count rate in the 25--300 keV band,  in units of
cnt cm$^{-2}$ s$^{-1}$.  }\\
\multicolumn{10}{l}{$^b$ Hardness ratio of the peak count rates in the 50--300 keV
to that in the 25--50 keV band. }\\
\multicolumn{10}{l}{$^c$ Coordinates. $^d$ Error radius. $^e$ Duration. }\\
\multicolumn{10}{l}{$^f$ Number of 1.024 s bins
where the signal exceeds 50 per cent of the peak value.}\\
\end{tabular}
\end{center}
\label{tab:21grbs}
\end{minipage}
\end{table*}

\begin{figure}
\centerline{\epsfig{file=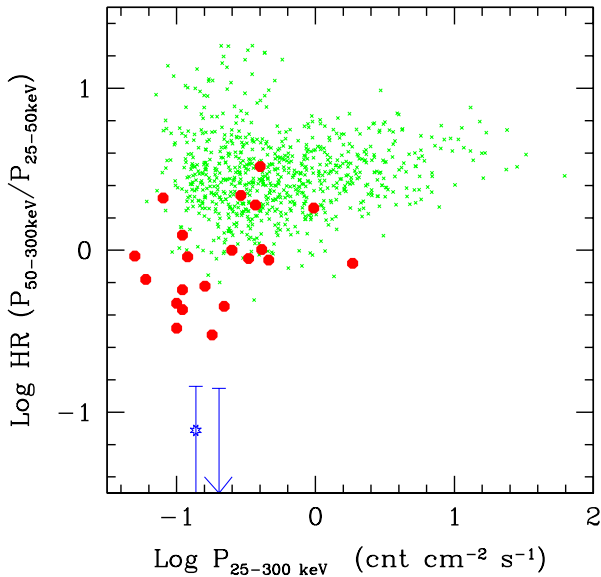,width=8.cm}}
\caption{Hardness-intensity distribution
of long ($>2$s) GRBs found in the \batse\ DISCLA records
 for the time period TJDs 11000-11699.
Hardness is estimated as the ratio of peak count rates
in 50--300 keV and   25--50 keV energy bands.
Peak count rate is given in 25--300 keV band.
GRBs from the scan of \citet{ste01} are marked by dots,
GRBs found in the present scan in 25--100 keV are shown
by circles, while two very soft events from the X-ray pulsar
Vela~X-1 are shown by symbols with errors bars.
}
\label{fig:hr_int}
\end{figure}

We present the results of our search for soft
\batse\ GRBs in Section~\ref{sec:search} and discuss new data
together with the recent GRB observations by
\sax\ and \hete\ in Section~\ref{sec:spectra}.

\section{Search for soft GRBs in the \batse\ records}
\label{sec:search}

We have performed the scan of
\batse\ DISCLA records
available at FTP archive at Goddart Space
Flight Center
\footnote{FTP archive at the Goddart Space
Flight Center is available at
ftp://cossc.gsfc.nasa.gov/compton/data/batse/daily }
for time period since  July 6, 1998
till  May 26, 2000 (TJDs  11000-11699, files for TJDs
11047, 11048, 11354, 11355-11359, 11519-11521
are missing).
The applied algorithm and technique is described
in \citet{ste01}. Only the 25--100 keV range
(1 and 2 channels) was checked.
The 1024 ms time resolution DISCLA data
are not suitable for studies of short ($<2$ s) GRBs
and we did not consider 1 bin events.
This allows us to avoid scintillation from
heavy nuclei and soft gamma-ray repeater (SGR) outbursts.
We excluded also events with location
in the vicinity of Galactic center, the
Sun, four known SGRs and other persistent sources
and events that  appeared
near and below Earth horizon.
We recorded only new GRBs missing in
the catalogs of \citet{pac99} and
\citet{ste01}.

We have found and classified as GRBs 21 new events.
Table~\ref{tab:21grbs} present their time identificator, intensity,
hardness, location and duration.
In the previous scan in 50--300 keV \citep{ste01}
for the same time period we have detected
about 800 long GRBs.
Hardness-intensity diagram (Fig.~\ref{fig:hr_int})
shows that although GRBs of a new sample
are softer on average, the samples do overlap.
Actually     13 out of 21 GRBs in a new sample
(Table~\ref{tab:21grbs}) and 23 of $\sim$800 long GRBs
in the old sample  \citep[][Table~\ref{tab:23softgrbs}]{ste01} have
    the peak count rate in 25--50 keV
higher than that in the 50--300 keV band.
According to this somewhat arbitrary criterion
we consider these 36 events as a sample
of   soft  long ($>$2 s) \batse\ GRBs.

\begin{table*}
\begin{minipage}{180mm}
\caption{Soft long ($>$2s) GRBs found by our previous scan
of the \batse\ DISCLA records (TJD~11000-11699) in the 50--300 keV band
\citep[][with location and duration data
from that catalog]{ste01}.
GRB980924, GRB981015 and GRB991006 are bright
and were first detected by the on-board procedure
\citep{pac99}.
GRB990304 was also detected by \konus\ and \ulysses,
GRB991217 and GRB000405 were observed by \ulysses.}
\begin{center}
\begin{tabular}{|rrrcrrrrrr|}
\hline
Date & Seconds & TJD$^a$ & $P^b$ & HR$^c$ & $\alpha^d$ & $\delta^d$ & $R^e$&  $T_{90}^f$& $N_{50}^g$\\
     &of TJD   &   &  &  & deg & deg & deg & s & \\
 \hline\hline
980924&  54262& 11080b& 0.95& 0.93&  61.8& $-$22.0&   8.8&  10&  2 \\
981015&  46766& 11101c& 1.36& 0.69& 122.9&  22.1&   5.4&  34&  6 \\
981115&  21438& 11132b& 0.80& 0.80& 284.0&  10.0&  10.6&   4&  1 \\
981117&  11629& 11134a& 0.41& 0.98& 217.6& $-$65.7&  23.7&   8&  3 \\
981118&   2533& 11135a& 0.48& 0.97& 186.9&  60.6&  23.0&   6&  2 \\
981128&  74360& 11145c& 0.18& 0.97&  60.3&  38.4&  33.2&  36&  3 \\
981204&  37850& 11151a& 0.16& 1.00&  53.4& $-$55.9&  21.5&   9&  3 \\
981222&  58180& 11169b& 0.30& 0.89& 145.9&  67.1&  28.7&  11&  5 \\
990112&   7066& 11190a& 0.20& 0.63& 118.6& $-$45.6&  17.7&  85& 14 \\
990207&  55697& 11216e& 0.49& 0.95& 152.9&  $-$9.7&  16.7&  17&  2 \\
990218&  73752& 11227b& 0.35& 0.98&  72.9&  37.7&  17.8&  89&  7 \\
990413&  32497& 11281d& 0.39& 0.94& 302.1&  55.5&  12.8&   6&  6 \\
990506&  42666& 11304c& 0.19& 0.89& 186.9&   9.6&  21.3&  61& 15 \\
990610&  56705& 11339c& 0.45& 0.99& 105.7& $-$16.6&   8.4& 109& 18 \\
990915&  58755& 11436c& 0.64& 0.78& 273.2& $-$21.9&   5.0&  50& 12 \\
991006&  68176& 11457b& 0.76& 0.99& 104.0&  11.7&   3.8&  73& 27 \\
991201&   1802& 11513a& 0.09& 0.98& 167.9& $-$10.9&  12.3&  19& 13 \\
991217&  37909& 11529b& 0.36& 0.49&  64.8& $-$12.7&  16.8&   8&  1 \\
991231&  28492& 11543a& 0.22& 0.96&  39.3&  32.3&  11.3&  14&  5 \\
000114&  51441& 11557a& 1.17& 0.99& 107.4& $-$25.3&   3.8&   5&  2 \\
000206&  74873& 11580g& 0.22& 0.98& 255.7&  78.5&  10.7&  26&  7 \\
000405&  77386& 11639b& 0.92& 0.96& 226.9& $-$52.5&   2.1&  35&  7 \\
000416&  52380& 11650c& 0.17& 0.70& 258.5& $-$65.7&  14.7&   9&  5 \\
\hline
\multicolumn{10}{l}{$^a$ Litera next to TJD means a name in the uniform catalog of
\citet{ste01}.}\\
\multicolumn{10}{l}{$^b$ Peak count rate in the 25--300 keV band,  in units of
cnt cm$^{-2}$ s$^{-1}$. }\\
\multicolumn{10}{l}{$^c$ Hardness ratio of the peak count rates in the 50--300 keV
to that in the 25--50 keV band.}\\
\multicolumn{10}{l}{$^d$ Coordinates. $^e$ Error radius. $^f$ Duration.}\\
\multicolumn{10}{l}{$^g$ Number of 1.024 s bins
where the signal exceeds 50 per cent of the peak value.}\\
\end{tabular}
\end{center}
\label{tab:23softgrbs}
\end{minipage}
\end{table*}

These 36 soft GRBs have typical light curves,
last up to about 100 s
and do not demonstrate any significant
anisotropy on the sky.
Soft \batse\ GRBs selected with the above criterion
constitute about 5\% of observed
long GRB sample (about 20 per year with peak fluxes
down to $\sim 0.1$ ph cm$^{-2}$ s$^{-1}$).

Our scan (as well as an alternative scan \citealt{kom01}) in
the \batse\ records in the 25--50 keV range  has yielded
surprisingly small number of new soft GRBs.
Moreover, there are no events with hardness ratio
(HR) below 0.3, while much softer events
like outbursts of Vela~X-1
can be confidently detected (see Fig.~\ref{fig:hr_int}).
We have considered sample of 50 events classified
by \citet{kom01} as unknown because of their softness.
When excluding short events they again have
$\mbox{HR}>0.3$ except two events
from the direction to Vela~X-1.

\begin{figure}
\centerline{\epsfig{file=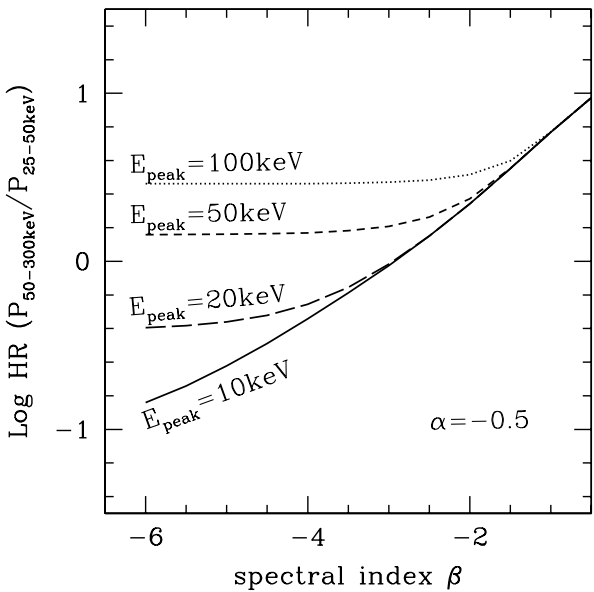,width=8.cm}}
\caption{Hardness ratio of GRBs
depending on different parameters
of the simulated incident photon spectra.
Spectra are approximated with the Band function
\citep{ban93} and folded with the \batse\
detector response matrix \citep{pen99}.
$\alpha$ and $\beta$ are low- and high-energy photon
spectral indices, respectively.}
\label{fig:hr_beta}
\end{figure}

Why do not we see softer GRBs despite the fact that there exist
X-ray dominated bursts with peak energy below the \batse\ window?
We calculated the \batse\ detector response
to estimate the dependence between the incident photon
spectrum and the observed hardness ratio (see Fig.~\ref{fig:hr_beta}).
We approximated GRB spectra with the Band function \citep{ban93}
and folded them with the \batse\
Detector Response Matrix (DRM) \citep{pen99}.
The low hardness ratio ($\mbox{HR}<1$) of our soft events is
consistent with a wide variety of spectral parameters
$E_{\rm peak}$, low energy and high energy spectral
indices $\alpha$ and $\beta$, in particular,
with $E_{\rm peak} < 20$ keV and $\beta < -3$ (see Fig.~\ref{fig:hr_int}
and \ref{fig:hr_beta}).
It is also evident  that the sufficient condition for a GRB to
give $\mbox{HR}>1$ and thus to look as a typical GRB in the \batse\
data (with larger signal above 50 keV) is $\beta > -3$,
independently on the $E_{\rm peak}$.
A combination of a low $E_{\rm peak}$ and a very steep $\beta$
would give the hardness ratio below 0.3 which we do not observe.
The fact that all events with $\mbox{HR}<0.3$ have
evident non-GRB origin (solar flares, Vela~X-1 pulsar, etc.) implies
that a spectral cutoff below $\sim$15 keV
may be a distinguishing feature to separate non-GRB sources.

\section{GRB spectra as observed by \batse/\sax/\hete}
\label{sec:spectra}

\begin{figure*}
\centerline{\epsfig{file=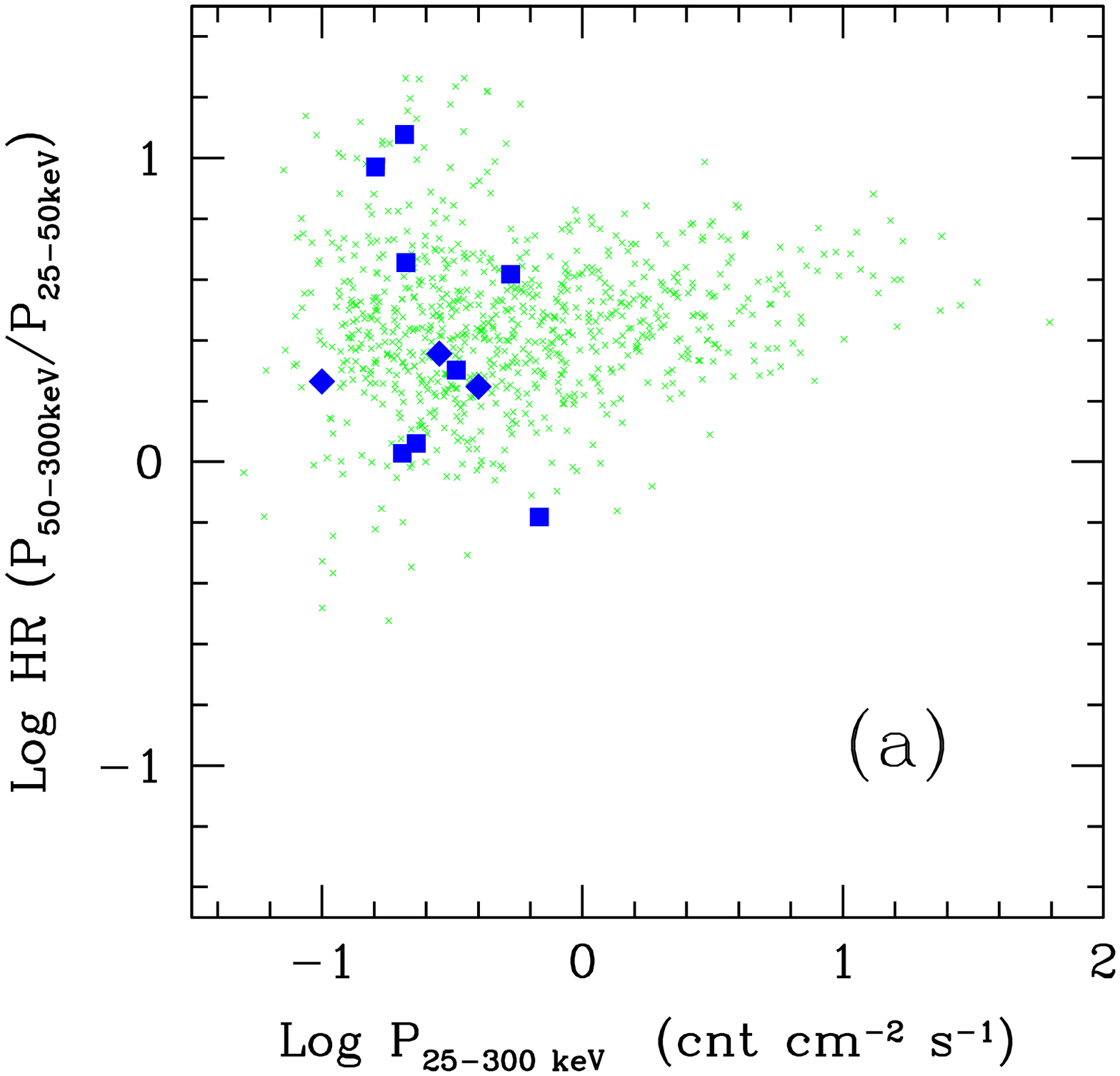,width=8.cm}
\epsfig{file=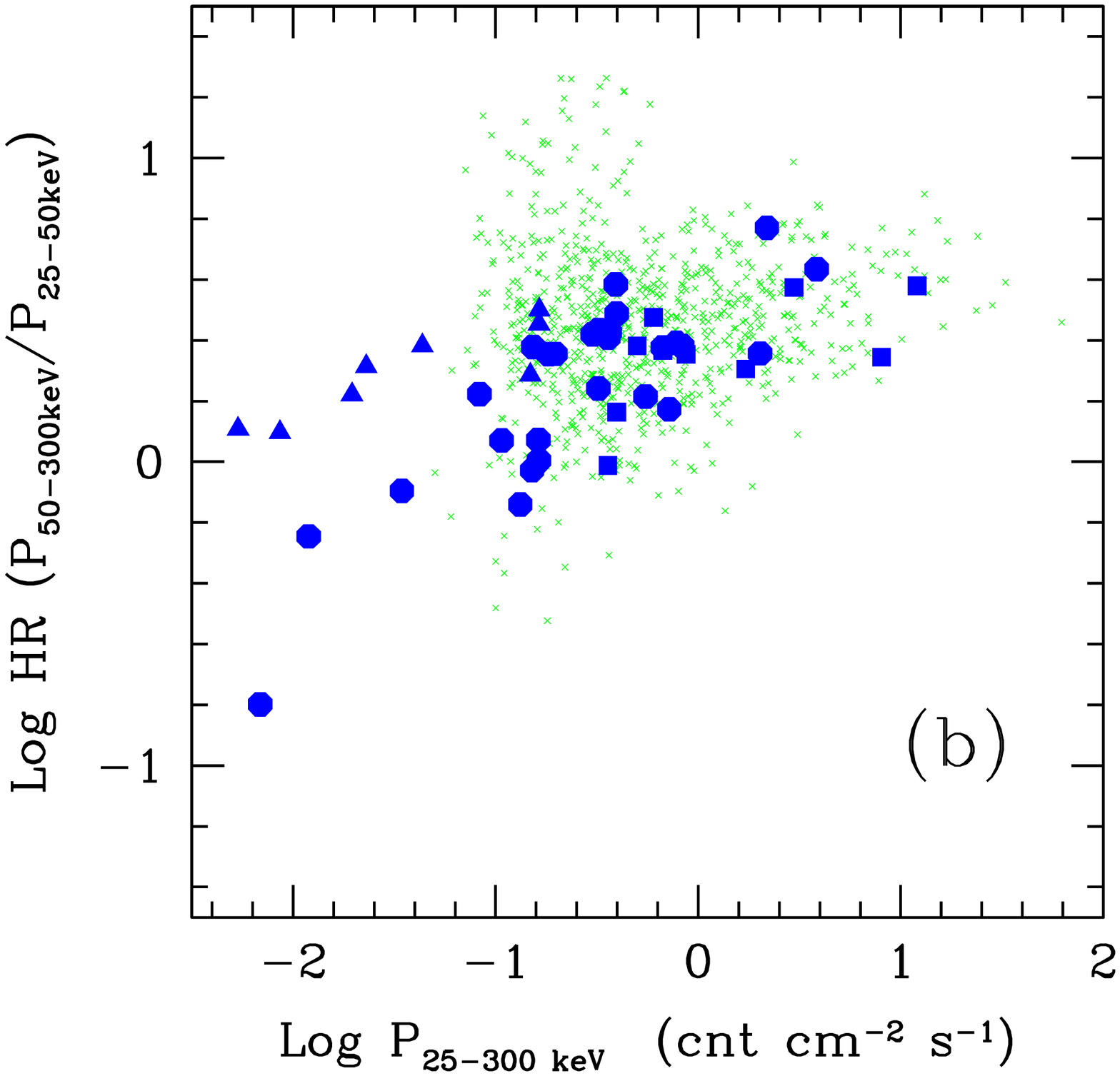,width=8.cm}}
\caption{
(a) Hardness-intensity distribution of
of GRBs detected by \sax\ as seeing in the \batse\ data.
Squares show the events from
\citet{kip01} and diamonds are the
events reported in \citet{intz03}.
(b) Hardness-intensity distribution
of the model \batse\ counterparts
of GRBs detected by \hete\ \citep{sak04}.
Circles show the best-fitting result with the
power-law with the exponential cutoff spectrum,
triangles correspond to the best-fitting
power-law spectrum, while squares  are for the
fits with the Band function.
In both panels, the  dots represent   the \batse\ GRB sample
from the time interval TJD 11000--11699.
}
\label{fig:saxhete}
\end{figure*}

\sax\ observed 20 X-ray dominated GRBs which were
detected by Wide Field Camera (2--26 keV), but did not
activate the trigger of the Gamma-Ray Monitor (40--400 keV).
Their counterparts  were found in the \batse\ records
for almost all observable events \citep{kip01,intz03}.
Most of them were detected earlier
as classic GRBs by our scan of \batse\ data
in 50--300 keV band \citep{ste01}.
It turns out that these events have a high hardness ratio similar to
typical GRBs (see Fig.~\ref{fig:saxhete}a).
The hardness ratio 100--300/50--100 keV for common \sax/BATSE events shows
a similar picture: 7 out of 8 events have a typical hardness for weak GRBs
and one event is softer. Thus this distribution is also
consistent with the extrapolated hardness-intensity trend for
long GRBs \citep{kip01}.
These facts support our conclusion that
most of the X-ray dominated GRBs should have a hard tail with $\beta > -3$
in the \batse\ window 25--1000 keV (see Fig.~\ref{fig:hr_beta}).

\hete\ observed 45 GRBs in the 2--400 keV band and
their spectral fits are given in \citet{sak04}.
In order to check how \hete\ results are related to our data
we folded \hete\  spectra with the \batse\ detector response matrix
and obtained corresponding counts in \batse\ channels
(see Fig.~\ref{fig:saxhete}b).
The fraction of soft events in the \hete\  sample
is about 3 times larger than in the \batse\ sample which
can be explained by  different instrument responses.
But only 1 out of 45 \hete\ events gives a lower hardness
ratio than we see in the \batse\ GRB sample.
Nine  out of 45 \hete\ events are below the \batse\ sensitivity threshold.
The \batse\ sample, however, represents probably
the whole GRB spectral variety.

\citet{sak04} fitted \hete\ spectra by three functions:
a power-law, a power-law with the exponential cutoff
and the Band function. They started from
a power-law fit and used a more parametric expression only
if the fit was inconsistent with the data at 0.01 significance level.
From Fig.~\ref{fig:saxhete}b
one can see that the power-law fit was acceptable only for
weak events. Relatively bright bursts gave good
power-law with exponential cutoff fits.
However, this does not mean that they could not
be fitted by the Band function.
 According to our results, the existence of GRBs with sharp spectral
cutoff is questionable for events with low $E_{\rm peak}$. Indeed,
the events with $E_{\rm peak} \sim$10--20 keV would give a very low
hardness ratio which we do not observe.
Note that, as shown by \citet{pre00}, only few percents of
GRBs with high $E_{\rm peak}$ are better described by
a power-law  with the exponential cutoff.
If the dispersion in $E_{\rm peak}$ is due to
variations in the redshift/blueshift in the source,
then the spectral shape would be stable and
our conclusion could refer to all GRBs.

\section{Conclusions}

\begin{enumerate}

 \item Despite the wealth of the X-ray dominated GRBs
observed by \ginga, \sax\ and \hete\
the number of soft GRBs in the \batse\ data is
relatively small. The fraction of events with
the count rate in 25--50 keV higher
than that above 50 keV is $\sim$5 per cent
(20 per year with flux down to 0.1 ph cm$^{-2}$ s$^{-1}$).

\item The hardness distribution of the X-ray dominated GRBs in the
BATSE  band is consistent with that of weak classic GRBs.
In the case of a low $E_{\rm peak}$, the main fraction of GRBs should have a
relatively hard high-energy tail with a power-law slope $\beta >-3$.
Only a few per cent of the X-ray rich GRBs have a tail with
$\beta <-3$, but still harder than the exponential one.
This fact clarifies the deficiency of soft events in the BATSE data.

 \item  An exponential cutoff in the GRB spectra,
if exists, is probably a  rare phenomenon.
Therefore, a spectral cutoff with the e-folding energy below
$\sim 20$ keV may be a distinguishing feature to
separate the non-GRB events.
\end{enumerate}

\section*{Acknowledgments}

We are grateful to Robert Preece and Geoffrey Pendleton
for the code of the \batse\ detector response matrix.
We thank Kevin Hurley for providing us \ipn\ data on
our soft GRB sample.
This research has made use of data obtained through
the High Energy Astrophysics Science Archive Research Center Online Service,
provided by the NASA/Goddard Space Flight Center.
The work is supported by NORDITA Nordic project
in high energy astrophysics in the INTEGRAL era,
Russian Foundation for Basic Research (grant 04-02-16987),
and Russian Foundation of Science Support (Y.T.).

\label{lastpage}

\end{document}